\documentclass{PoS}

\usepackage{amsmath}
\usepackage{graphicx}

\title{Hyperon-nucleon Scattering In A Covariant Chiral Effective Field Theory Approach}

\ShortTitle{Hyperon-nucleon Scattering In A Covariant Chiral Effective Field Theory Approach}

\author{\speaker{Kai-Wen Li}%
         \\
        School of Physics and Nuclear Energy Engineering and International Research Center for Nuclei and Particles in the Cosmos, Beihang University, Beijing 100191, China\\
        E-mail: \email{kaiwen.li@buaa.edu.cn}}

\author{Xiu-Lei Ren\\
        School of Physics and State Key Laboratory of Nuclear Physics and Technology, Peking University, Beijing 100871, China\\
        E-mail: \email{xiulei.ren@pku.edu.cn}}

\author{Li-Sheng Geng\\
        School of Physics and Nuclear Energy Engineering and International Research Center for Nuclei and Particles in the Cosmos, Beihang University, Beijing 100191, China\\
        Beijing Key Laboratory of Advanced Nuclear Materials and Physics, Beihang University, Beijing 100191, China\\
        E-mail: \email{lisheng.geng@buaa.edu.cn}}

\author{Bingwei Long\\
        Center for Theoretical Physics, Department of Physics, Sichuan University, 29 Wang-Jiang Road, Chengdu, Sichuan 610064, China\\
        E-mail: \email{bingwei@scu.edu.cn}}

\abstract{A recently proposed covariant chiral effective field theory approach is applied to study strangeness $S=-1$ hyperon-nucleon interactions at leading order. 12 low energy constants are introduced by Lorentz invariance, which is different from the heavy baryon approach, where only five appear. The Kadyshevsky equation is employed to iterate the chiral potentials. A quite satisfactory description of the 36 hyperon-nucleon scattering data is obtained with $\chi^2\simeq 17$, which is comparable with the next-to-leading order heavy baryon approach. The results hint at a more efficient way to construct the chiral potentials.}

\FullConference{The 26th International Nuclear Physics Conference\\
		 11-16 September, 2016\\
		 Adelaide, Australia}

\begin{document}

\section{Introduction}

Baryon-baryon interactions play a crucial role in hypernuclear physics. 
During the past 20 years, chiral effective field theory ($\chi$EFT),  first proposed by Weinberg~\cite{Weinberg:1990rz, Weinberg:1991um},   has made great progress in deriving nucleon-nucleon ($NN$)~\cite{Bedaque:2002mn, Epelbaum:2008ga, Machleidt:2011zz}, hyperon-nucleon ($YN$)~\cite{Polinder:2006zh, Haidenbauer:2013oca}, hyperon-hyperon~\cite{Polinder:2007mp, Haidenbauer:2009qn, Haidenbauer:2015zqb} interactions.

Nevertheless, current $\chi$EFT studies on baryon-baryon interactions, referred to as the heavy baryon (HB) approach, are based on a non-relativistic theoretical framework, which cannot be used for relativistic hypernuclear few- and many-body calculations. In addition, the HB approach is very  sensitive to ultraviolet cutoffs, known as the renormalization group invariance problem~\cite{Lepage:1997cs,Long:2007vp}. Recently, Epelbaum and Gegelia proposed a new $\chi$EFT approach~\cite{Epelbaum:2012ua,Epelbaum:2015sha} (referred to as the EG approach) for the $NN$ interaction. At leading order (LO), the potential is the same as that of the HB approach but the scattering equation changes to the Kadyshevsky equation. It is shown that in the strangeness $S=-1$ $YN$ system~\cite{Li:2016paq}, the description of the scattering data with the EG approach remains quantitatively similar to the HB approach, only the cutoff dependence is mitigated.

A new covariant power counting is recently proposed to study the $NN$~\cite{Ren:2016jna} and $YN$~\cite{Li:2016mln} systems in $\chi$EFT, which is partly motivated by the successes of covariant chiral perturbation theory in heavy-light and one baryon systems~\cite{Geng:2008mf,Ren:2012aj,Geng:2010df,Geng:2013xn}. At the potential level, the baryons are treated in a covariant way to maintain all the symmetries and analyticities. In the present work, we report on the study of the strangeness $S=-1$ $\Lambda N-\Sigma N$ system in this covariant $\chi$EFT approach at LO.

\section{Formalism}

\begin{figure}
  \centering
  \includegraphics[width=0.105\textwidth]{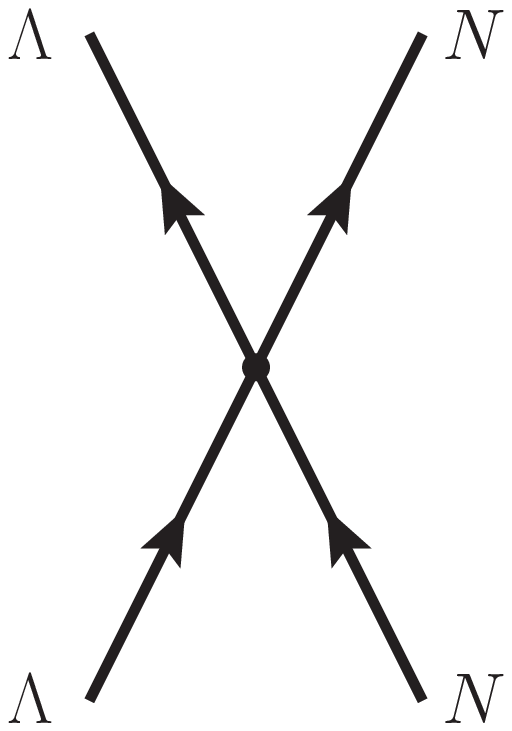}
  \includegraphics[width=0.105\textwidth]{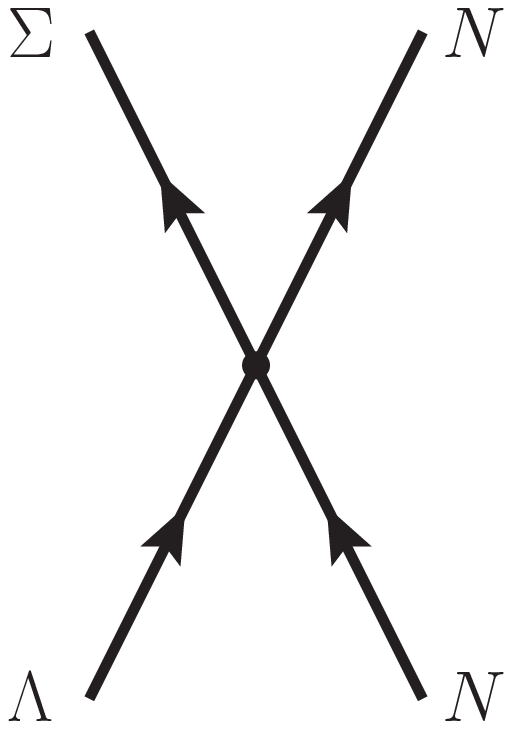}
  \includegraphics[width=0.105\textwidth]{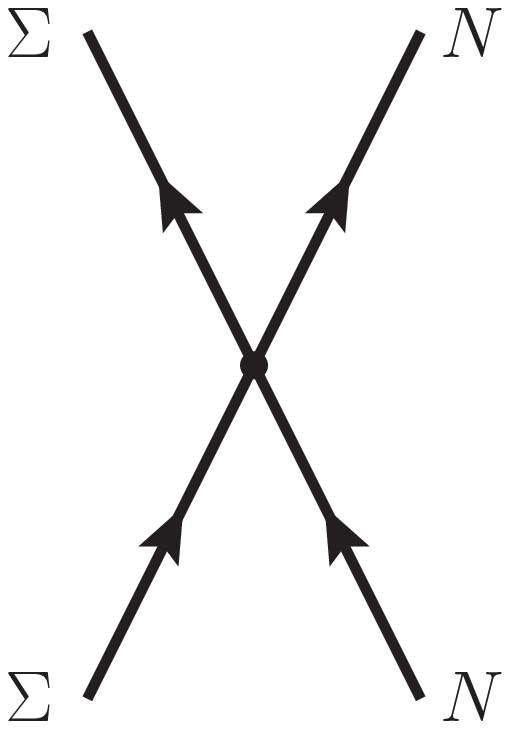}
  \includegraphics[width=0.105\textwidth]{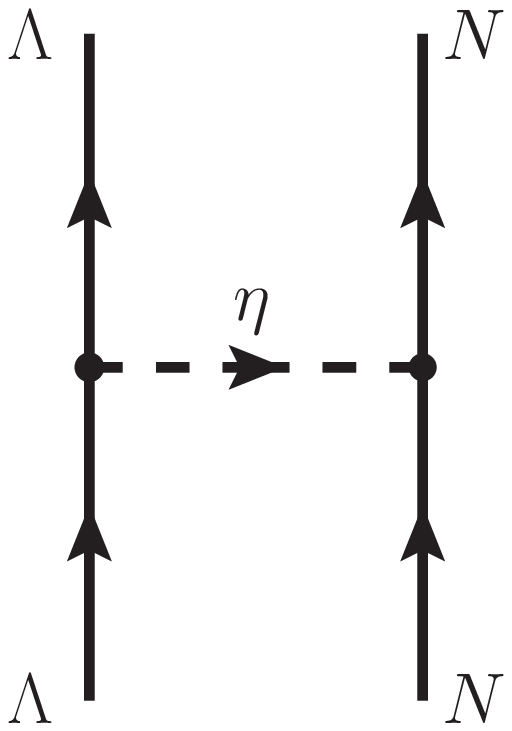}
  \includegraphics[width=0.105\textwidth]{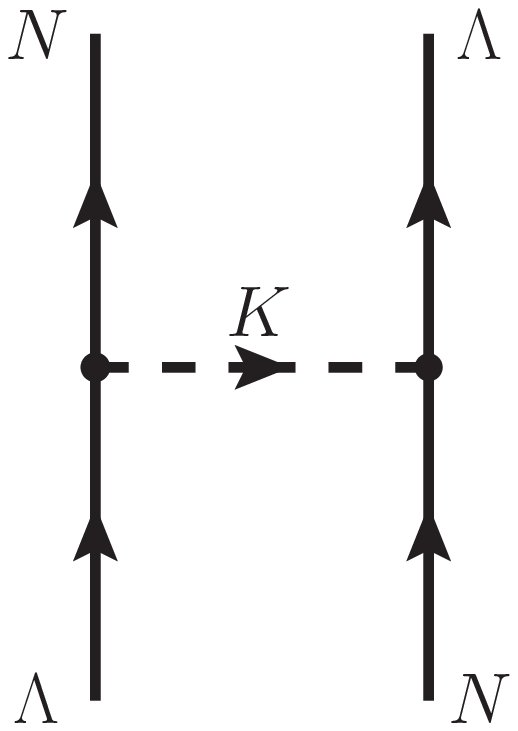}
  \includegraphics[width=0.105\textwidth]{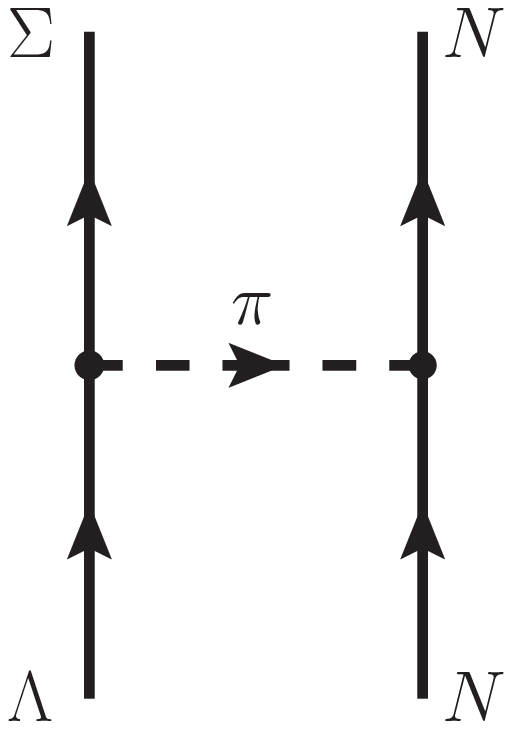}
  \includegraphics[width=0.105\textwidth]{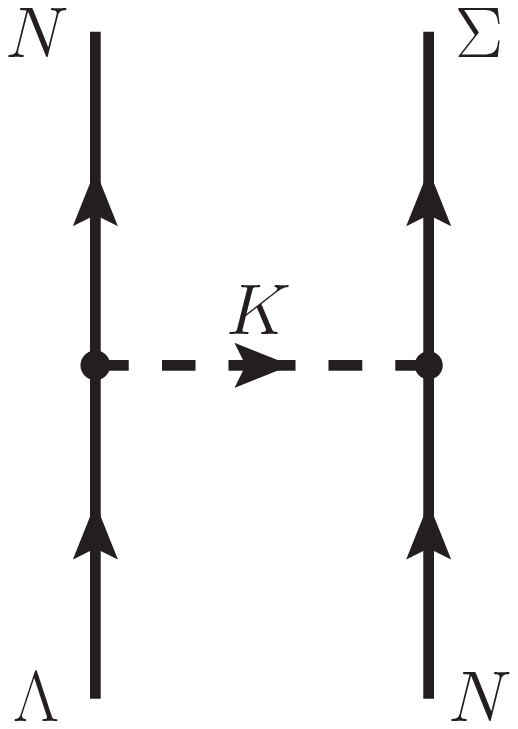}
  \includegraphics[width=0.105\textwidth]{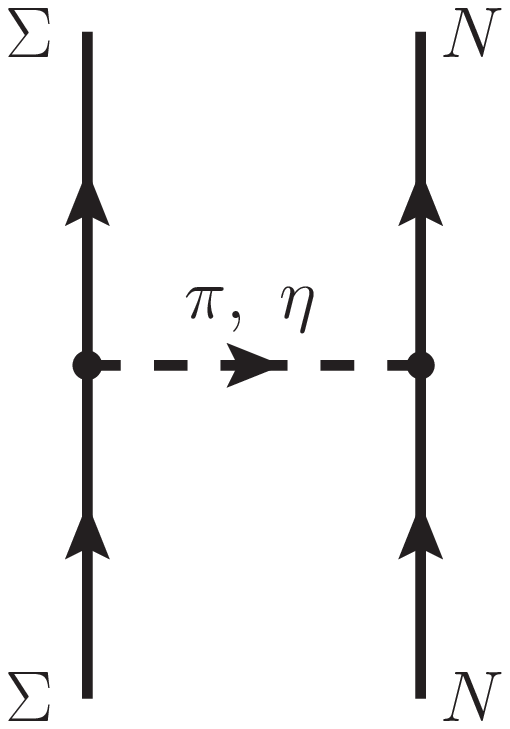}
  \includegraphics[width=0.105\textwidth]{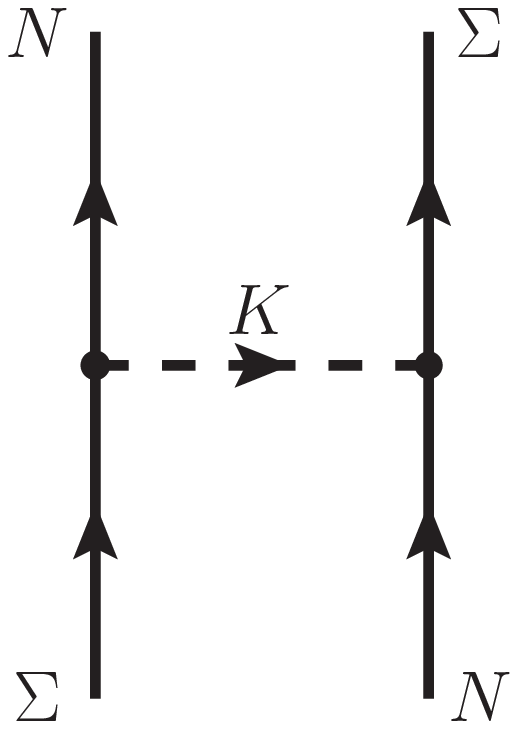}
  \caption{Nonderivative four-baryon contact terms and one-pseudoscalar-meson-exchange diagrams for the $\Lambda N-\Sigma N$ system. The solid lines denote incoming and outgoing baryons, and the dashed lines denote the exchanged pseudoscalar mesons.}\label{CTOME1}
\end{figure}

The details can be found in Refs.~\cite{Ren:2016jna,Li:2016mln}.  In the following we briefly explain the procedures in obtaining the chiral potentials, calculating the scattering equation, and performing the fits. We start from the complete Dirac spinor so that Lorentz invariance is retained,
\begin{equation}
  u_B(\mbox{\boldmath $p$}, s)= N_p
  \left(
  \begin{array}{c}
    1 \\
    \frac{\mbox{\boldmath $\sigma$}\cdot \mbox{\boldmath $p$}}{\epsilon_B}
  \end{array}\right)
  \chi_s,
  ~~~~\qquad\qquad
  N_p=\sqrt{\frac{\epsilon_B}{2M_B}},
\end{equation}
where $\epsilon_B=E_B+M_B$ and $E_B=\sqrt{\mbox{\boldmath $p$}^2+M_B^2}$, while the small component of $u_B$ is omitted by employing a non-relativistic reduction in the HB approach. Here naive dimensional analysis is applied in obtaining the Feynman diagrams at LO, which consist of nonderivative four-baryon contact terms and one-pseudoscalar-meson exchanges. Specifically, following Refs.~\cite{Polinder:2006zh, Li:2016paq}, there are 3 for the former and 7 for the latter in the strangeness $S=-1$ $\Lambda N-\Sigma N$ system, as shown in Figure.~\ref{CTOME1}.

The relevant chiral Lagrangians in this framework consist of the nonderivative four-baryon contact terms and the meson-baryon interactions

\begin{align}
  &\mathcal{L}_{\textrm{CT}}^1 = \frac{C_i^1}{2}~\textrm{tr}\left(\bar B_a \bar B_b (\Gamma_i B)_b (\Gamma_i B)_a\right)\, ,\qquad\qquad
  \mathcal{L}_{\textrm{CT}}^2 = \frac{C_i^2}{2}~\textrm{tr}\left(\bar B_a (\Gamma_i B)_a \bar B_b (\Gamma_i B)_b\right)\, , \nonumber\\
  &\mathcal{L}_{\textrm{CT}}^3 = \frac{C_i^3}{2}~\textrm{tr}\left(\bar B_a (\Gamma_i B)_a\right)\textrm{tr}\left( \bar B_b (\Gamma_i B)_b\right), \\
  &\mathcal{L}_{\textrm{MB}}^{(1)} =
  \mathrm{tr}\Bigg( \bar B \big(i\gamma_\mu D^\mu - M_B \big)B -\frac{D}{2} \bar B \gamma^\mu\gamma_5\{u_\mu,B\}
  -  \frac{F}{2}\bar{B} \gamma_\mu\gamma_5 [u_\mu,B]\Bigg)\, ,
\end{align}
where $\mathrm{tr}$ indicates trace in flavor space; $\Gamma_i$ are the elements of the Clifford algebra,
and $C_i^m$ ($m=1,2,3$) are the low energy constants (LECs) corresponding to independent four-baryon operators. In addition, $D^\mu B = \partial_\mu B+[\Gamma_\mu,B]$, $\Gamma_\mu = \frac{1}{2}\left[u^\dag\partial_\mu u + u\partial_\mu u^\dag \right]$, $u_\mu=i(u^\dagger \partial_\mu u-u\partial_\mu u^\dagger)$ and $u^2= U = \exp\left(i\frac{\sqrt{2}\phi}{f_0}\right)$, where $f_0$ is the pseudoscalar meson decay constant in the chiral limit. The matrices $B$ and $\phi$
collect the octet baryons and pseudoscalar mesons. In numerical calculations, we use $D+F =1.26$, $F/(F+D)=0.4$ and $f_0=93$ MeV~\cite{Haidenbauer:2013oca}. Potentials for the Feynman diagrams in Figure.~\ref{CTOME1} are derived from the above Lagrangians. Strict SU(3) symmetry is imposed on the contact terms thus  12 independent LECs remain: $C_{1S0}^{\Lambda\Lambda}$, $\hat C_{1S0}^{\Lambda\Lambda}$, $C_{1S0}^{\Sigma\Sigma}$, $\hat C_{1S0}^{\Sigma\Sigma}$, $C_{3S1}^{\Lambda\Lambda}$, $\hat C_{3S1}^{\Lambda\Lambda}$, $C_{3S1}^{\Sigma\Sigma}$, $\hat C_{3S1}^{\Sigma\Sigma}$, $C_{3S1}^{\Lambda\Sigma}$, $\hat C_{3S1}^{\Lambda\Sigma}$, $C_{3P1}^{\Lambda\Lambda}$, $C_{3P1}^{\Sigma\Sigma}$. SU(3) symmetry is only broken by the mass difference of the mesons in the one-pseudoscalar-meson exchanges.  The potentials corresponding to the Feynmann diagrams  shown in Fig.~1 can be expressed in the following way: \begin{align}
  V^{B_1B_2\rightarrow B_3B_4}_{\textrm{CT}} = C_i\left(\bar u_3 \Gamma_i u_1\right)\left(\bar u_4 \Gamma_i u_2\right),
\end{align}
\begin{equation}\label{VOMEP}
V^{B_1B_2\rightarrow B_3B_4}_\mathrm{OME} = -N_{B_1B_3\phi}N_{B_2B_4\phi}
   \frac{(\bar u_3 \gamma^\mu \gamma_5 q_\mu u_1) (\bar u_4 \gamma^\nu \gamma_5 q_\nu u_2)}
        {q_0^2-q^2-m^2}\mathcal{I}_{B_1B_2\rightarrow B_3B_4}\, ,
\end{equation}
where the subscripts CT and OME denote the contact term and one-pseudoscalar-meson exchange, respectively. $N_{BB\phi}$ is the SU(3) coefficient and $\mathcal{I}_{B_1B_2\rightarrow B_3B_4}$ the isospin factor. Ddetails can be found in Refs.~\cite{Polinder:2006zh,Ren:2016jna,Li:2016mln}.

The Kadyshevsky equation is used to iterate the potentials due to the infrared enhancement in two-baryon propagations following Ref.~\cite{Li:2016paq}:
\begin{align}\label{SEK}
  & T_{\rho\rho'}^{\nu\nu',J}(\mbox{\boldmath $p$}',\mbox{\boldmath $p$};\sqrt{s})
  =
   V_{\rho\rho'}^{\nu\nu',J}(\mbox{\boldmath $p$}',\mbox{\boldmath $p$})
    +
  \sum_{\rho'',\nu''}\int_0^\infty \frac{dp''p''^2}{(2\pi)^3} \frac{2\mu_{\nu''}^2~ V_{\rho\rho''}^{\nu\nu'',J}(\mbox{\boldmath $p$}',\mbox{\boldmath $p$}'')~
   T_{\rho''\rho'}^{\nu''\nu',J}(\mbox{\boldmath $p$}'',\mbox{\boldmath $p$};\sqrt{s})}{\left(\mbox{\boldmath $p$}''^2+4\mu_{\nu''}^2\right)
  \left(\sqrt{\mbox{\boldmath $q$}_{\nu''}^2+4\mu_{\nu''}^2}-\sqrt{\mbox{\boldmath $p$}''^2+4\mu_{\nu''}^2}+i\epsilon\right)},
\end{align}
where $\sqrt{s}$ is the total energy of the baryon-baryon system in the center-of-mass frame, $\mbox{\boldmath $q$}_{\nu''}$ is the relativistic on-shell momentum. The labels $\rho$ and  $\nu$ denote the partial waves and particle channels, respectively.
Relativistic kinematics is used for the relation of the laboratory momenta and the center-of-mass momenta. We solve the Kadyshevsky equation in particle basis in order to account for the physical thresholds properly. The Coulomb force is treated with the Vincent-Phatak method~\cite{Vincent:1974zz}. To regularize the integration in the high-momentum region, the chiral potentials are multiplied with an exponential form factor
$
  f_{\Lambda_F}(\mbox{\boldmath $p$},\mbox{\boldmath $p$}') = \exp \left[-\left(\mbox{\boldmath $p$}/\Lambda_F\right)^{4}-\left(\mbox{\boldmath $p$}'/\Lambda_F\right)^{4}\right].
$
One should note that although the potentials are fully covariant, the scattering amplitude is not because of the use of the Kadyshevsky equation and the cutoff regularization.

The LECs can be pinned down by fitting to the 36 $YN$ scattering data, as it has been done in Ref.~\cite{Haidenbauer:2013oca}. In actual calculations we introduce the following constraints, i.e. the $\Lambda p$ $^1S_0$, $^3S_1$ and $\Sigma^+ p$ $^3S_1$ scattering lengths. The $\Lambda$ hypertriton is not taken into account because we are not able to perform 3-body calculations at present.

\section{Results and discussion}
\begin{figure}
  \centering
  \includegraphics[width=0.7\textwidth]{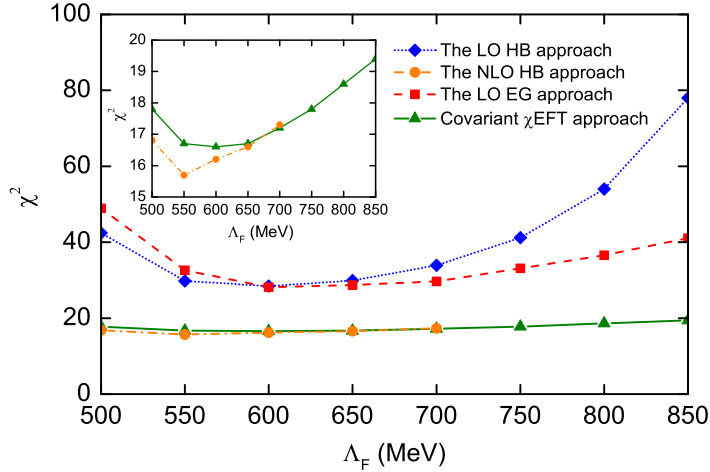}
  \caption{$\chi^2$ as a function of the cutoff in the LO (blue dotted line)~\cite{Polinder:2006zh}, NLO (orange dashed-dotted line)~\cite{Haidenbauer:2013oca} HB approach, the LO EG approach (red dashed line)~\cite{Li:2016paq} and the LO covariant $\chi$EFT approach (green solid line).}\label{chi2}
\end{figure}

\begin{figure}
  \centering
  \includegraphics[width=1\textwidth]{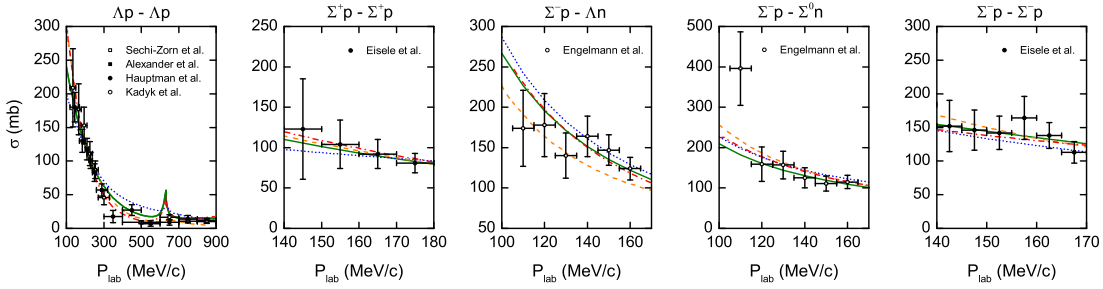}
  \caption{Cross sections in the HB approach (blue dotted lines)~\cite{Polinder:2006zh} and covariant $\chi$EFT approach (green solid lines) at LO as functions of the laboratory momentum at $\Lambda_F=600$ MeV. For reference, the NSC97f~\cite{Rijken:1998yy} (red dash-dotted lines) and J\"ulich 04~\cite{Haidenbauer:2005zh} (orange dashed lines) results are also shown. The experimental data are the same as Ref.~\cite{Li:2016paq}.}\label{cs}
\end{figure}

The fit was done by varying the cutoff value $\Lambda_F$ around the $\rho$ meson mass. Cutoff dependence of the $\chi^2$ in describing the 36 $YN$ data  is shown in Figure.~\ref{chi2},  in comparison with other $\chi$EFT approaches. It can be seen that the LO covariant $\chi$EFT approach yields a smaller $\chi^2$ value and
a milder cutoff dependence of the $\chi^2$, compared with the LO HB and EG approaches. They are similar to the next-to-leading (NLO) HB approach, which has 23 LECs, however. Nevertheless, renormalization group invariance is still not achieved, as exhibited by the variation of $\chi^2$ with the cutoff (more clearly shown in the insert).
The minimum value of the $\chi^2$ is 16.7, located at $\Lambda_F=600$ MeV, and the corresponding LECs are listed in Table~\ref{tab_LECs600}.
\begin{table}
\centering
 \caption{Low-energy constants (in units of $10^4$ GeV$^{-2}$) at $\Lambda_F=600$ MeV in the covariant $\chi$EFT approach.}
\renewcommand\arraystretch{1}
   \begin{tabular}{ccccccc}
  \hline
  \hline
  LECs & $C^{\Lambda \Lambda}_{1S0}$ & $C^{\Sigma \Sigma}_{1S0}$ & $C^{\Lambda \Lambda}_{3S1}$ & $C^{\Sigma \Sigma}_{3S1}$ & $C^{\Lambda \Sigma}_{3S1}$ & $\hat C^{\Lambda \Lambda}_{1S0}$ \\
  & $-0.0223$ & $-0.0528$ & $0.0032$ & $0.0842$ & $0.0232$ & $3.9185$ \\
  \hline
  LECs & $\hat C^{\Sigma \Sigma}_{1S0}$ & $\hat C^{\Lambda \Lambda}_{3S1}$ & $\hat C^{\Sigma \Sigma}_{3S1}$ & $\hat C^{\Lambda \Sigma}_{3S1}$ & $C^{\Lambda \Lambda}_{3P1}$ & $C^{\Sigma \Sigma}_{3P1}$ \\
     & $4.0681$ & $0.4190$ & $-0.4132$ & $0.7326$ & $0.2044$ & $0.2616$ \\

  \hline
  \hline
\end{tabular}\label{tab_LECs600}
 \end{table}
Note that the LECs in the $\Lambda p$ $^1S_0$ partial wave cannot be precisely determined for the $\Lambda$ hypertriton binding energy is not used as a constraint. We show only a typical case here. Moreover, the description of the experimental data  are shown in Fig.~\ref{cs}.
The LO HB approach~\cite{Polinder:2006zh} and two phenomenological models~\cite{Rijken:1998yy,Haidenbauer:2005zh} are taken as references. Clearly the covariant $\chi$EFT approach can better reproduce the data compared with the LO HB approach, especially for the $\Lambda p\rightarrow\Lambda p$ and $\Sigma^+ p\rightarrow \Sigma^+ p$ reactions.

The improvements in our new scheme mainly come from the contact terms. In the LO HB approach, contact terms only contribute to central and spin-spin potentials, and they only appear in $S$-waves. In our new scheme, tensor, spin-orbit, asymmetric spin-orbit and quadratic spin-orbit terms also appear in the contact interactions. These terms are momentum dependent and  improve the description. These are parallel to the $NN$ case. Detailed discussions can be found in Ref.~\cite{Ren:2016jna}.

\section{Summary}

We have applied a new covariant chiral effective field theory approach to the strangeness $S=-1$ $\Lambda N-\Sigma N$ system at leading order. Starting from the covariant chiral Lagrangians, the full baryon spinor is retained to calculate the potentials. As such, Lorentz invariance is preserved. We use strict SU(3) symmetry in the contact terms thus only 12 low energy constants are independent. SU(3) symmetry is broken in the one-pseudoscalar-meson exchanges due to the mass difference of the exchanged mesons. The Kadyshevsky equation is employed to iterate the chiral potentials. A very satisfactory description of the 36 hyperon-nucleon scattering data is obtained, as well as a mitigated cutoff dependence. Our approach yield results similar to the next-to-leading order heavy baryon approach but with 11 fewer low energy constants. It hints at a more efficient way to construct the baryon-baryon potentials.

\section*{Acknowledgements}

This work is partly supported by the National Natural Science Foundation of China under Grants No. 11375024, No. 11522539, and No. 11375120, the China Postdoctoral Science Foundation under Grant No. 2016M600845, and the Fundamental Research Funds for the Central Universities.

\end{document}